# Study of vortex dynamics in an *a*-MoGe thin film using low-frequency two-coil mutual inductance measurements

Soumyajit Mandal[1], Somak Basistha, John Jesudasan, Vivas Bagwe, and Pratap Raychaudhuri

*Tata Institute of Fundamental Research, Homi Bhabha Road, Mumbai 400005.*

**Abstract**

We extract the vortex lattice parameters using low-frequency two-coil mutual inductance measurements in a 20-nm-thick superconducting *a*-MoGe thin film. We fit the temperature dependence of ac penetration depth in the mixed state using a model developed by Coffey and Clem and demonstrate a procedure for extracting vortex lattice parameters such as pinning constant, vortex lattice drag coefficient, and pinning potential barrier. We show that the extracted parameters follow the magnetic field variation expected for a weakly pinned 2-dimensional vortex lattice.

[1] Email: soumyajitmandal13@gmail.com

# 1. Introduction

Understanding the dynamics of vortices in type II superconductors is of paramount importance both from a fundamental standpoint and for the practical application of these materials[1, 2, 3, 4, 5, 6]. In type II superconductors, when we apply a magnetic field larger than the lower critical field $H_{c1}$, quantized flux lines (vortices) penetrate the sample. In a clean superconductor, the interaction between the vortices arranges them in a triangular lattice known as the Abrikosov[7] vortex lattice (VL). However, the inevitable presence of crystalline defects in solid acts as a random pinning potential for the vortices. If these vortices are made to oscillate under the influence of an oscillatory current or magnetic field, their motion is governed by the following competing forces[8]: i) Lorentz force due to the external current density driving the motion, ii) restoring force due to the combined effect of pinning by crystalline defects and repulsion from neighbouring vortices, and iii) the dissipative viscous drag of the vortices. In addition, at finite temperature thermal activation can cause the vortices to spontaneously jump over the pinning barrier resulting in thermally activated flux flow (TAFF), which produces a small resistance even for external current much below the critical current density ($J_c$).

To model this problem, Gittleman and Rosenblum[9] (GR) considered the following simplified picture for a single vortex (neglecting vortex mass term and TAFF), where the displacement $\vec{u}$ of the vortex due to small ac excitation follows the following equation of motion (force per unit length of the vortex):

$$\eta \dot{\vec{u}} + \alpha_L \vec{u} = \phi_0 \vec{J_{ac}} \times \hat{n}, \qquad (1)$$

Where $\vec{J_{ac}}$ is the external alternating current density on the vortex containing one flux quantum $\phi_0$ ($\hat{n}$ being the unit vector along the vortex), $\eta$ is the viscous drag coefficient on the vortex in absence of pinning and flux creep, and the restoring pinning force constant $\alpha_L$ is called the Labusch parameter[10]. Even though this model does not explicitly invoke the interaction between vortices, it nevertheless captures the dynamics of a vortex solid where the pinning acts collectively over a typical length scale over which the vortices maintain their positional order, namely the Larkin length. In this case the resultant pinning parameters have to be interpreted in a mean-field sense[8], where they incorporate both the effect of interactions and the pinning potential. As we will show later, it can also be applied in certain vortex fluid states as long as the thermally activated motion of the vortices is very slow compared to the excitation frequency. $\alpha_L$ determines the shielding response of the superconductor in the vortex state. Assuming harmonic solution ($\sim e^{i\omega t}$) in the equation of motion (1), we get $\vec{u} = \phi_0 \frac{\vec{J_{ac}} \times \hat{n}}{(\alpha_L + i\omega\eta)}$. The resultant $\vec{u}$ modifies the London equation[11] in the following way[8]:

$$\vec{A} = -\mu_0 \lambda_L^2 \vec{J_{ac}} + \vec{u} \times \vec{B} = -\mu_0 \left( \lambda_L^2 + \frac{\phi_0 B}{\mu_0(\alpha_L + i\omega\eta)} \right) \vec{J_{ac}} = -\mu_0 \lambda_{eff}^2 \vec{J_{ac}}, \qquad (2)$$

Here, eqn. (2) has a form similar to the usual London equation, but the London[12] penetration depth ($\lambda_L$) is replaced by the effective complex penetration depth ($\lambda_{eff} = \sqrt{\left(\lambda_L^2 + \frac{\phi_0 B}{\mu_0(\alpha_L + i\omega\eta)}\right)}$). The Campbell penetration depth[13, 14, 15], $\lambda_C = \left(\frac{\phi_0 B}{\mu_0 \alpha_L}\right)^{1/2}$, is defined as the response from the vortices in the low-frequency limit, $\omega \ll \alpha_L/\eta$, when $\lambda_L$ can be ignored, i.e. $\lambda_L \ll \lambda_C$. Eqn. (2) captures the effect of small ac excitation on vortices, where $\lambda_{eff}$ is the effective ac screening length. On top of this, to account for the thermally activated flux motion over the pinning barriers, a random Langevin force which depends on the pinning barrier height $U$ is added to the right-hand side of eqn. (1). Physically, TAFF relaxes the restoring pinning force over large time scales and is therefore important when measurements are performed at very low frequencies. Consequently, the vortex state is characterized by the minimal set of three parameters: $\alpha_L, \eta$ and $U$.

In this paper, we demonstrate how the vortex lattice parameters can be extracted from $\lambda_{eff}$ measured using the two-coil mutual inductance technique[16, 17, 18, 19, 20] operating at tens of kHz frequencies. The sample under investigation is a very weakly pinned thin film of the amorphous superconductor MoGe. Here, the film is sandwiched between a primary drive coil and a secondary pickup coil, such that the ac magnetic field produced by the primary coil is partially shielded from the secondary, with the degree of shielding depending on the penetration depth of the superconductor. The advantage of this technique is that it allows precise determination of the absolute value of $\lambda_{eff}$, from the in-phase and out-of-phase mutual inductance of the two coils. We extract the vortex lattice parameters from the magnetic field and temperature dependence of $\lambda_{eff}$. Similar extraction of vortex parameters has been done before from the microwave resistivity[21]. However, in the microwave frequency range vortices get extremely small time to traverse during every single half-cycle of the ac drive which gives them very little room for thermal hopping. On the other hand, the effect of TAFF becomes more significant as the frequency is lowered making it possible to estimate both $U$ and $\alpha_L$ more accurately from our measurements.

## 2. Theoretical background

$\lambda_{eff}^2$ in eqn. (2) can be rewritten in terms of $\lambda_C$ or vortex resistivity, $\rho_v$, as:

$$\lambda_{eff}^2 = \lambda_L^2 + \lambda_C^2(1 + i\omega\tau_0)^{-1} = \lambda_L^2 - i\frac{\rho_v}{\mu_0\omega}, \tag{3}$$

where $\tau_0 = \eta/\alpha_L$ is the vortex relaxation time and $\rho_v$ is the complex vortex resistivity expressed in terms of dc flux flow resistivity ($\rho_{ff}$) as[22, 23]: $\rho_v = \rho_{ff}\frac{i\omega\tau_0}{1+i\omega\tau_0}$; $\rho_{ff} = \frac{B\phi_0}{\eta} = \frac{B}{B_{c2}}\rho_n$, where $\rho_n$ is the normal state resistivity. Looking at eqn. (3), we find two frequency regimes demarcated by the characteristic frequency, $\omega_0$ ($= 1/\tau_0$): The imaginary or the dissipative part becomes significant at a high enough frequency (flux-flow regime, $\omega > \omega_0$), while with decreasing frequency (Campbell regime, $\omega < \omega_0$), the real part dominates, and vortex contribution is given by $\lambda_C$.

So far, we have ignored TAFF. The effect of thermally activated flux jumps becomes important at very low frequencies or elevated temperatures. Typically, this is incorporated by assuming a typical single activation energy[8, 24], $U(T, H)$, in the same spirit that a single value of $\alpha_L$ was considered before. Thermally activated flux jump relaxes the restoring force on the vortices over a characteristic time scale $\tau$, which has been formally incorporated in slightly different ways by different authors[25, 26]. While Brandt[25] phenomenologically introduced the thermal creep by adding a temporal decay to the Labusch parameter $\alpha_L$, Coffey-Clem[26] (CC) did a slightly more elaborate calculation by adding a random thermal (Langevin) force term in the vortex equation of motion (1) to account for the TAFF motion and then solving it similar to the problem of a particle undergoing Brownian motion in a periodic potential[27, 28]. In their model, the expression of vortex resistivity ($\rho_v^{TAFF}$) gets modified to the following form[29]:

$$\rho_v^{TAFF} = \rho_{ff}\frac{\epsilon + i\omega\tau}{1 + i\omega\tau}, \tag{4}$$

where $\tau = \tau_0 \frac{I_0^2(\nu)-1}{I_1(\nu)I_0(\nu)}$ is the vortex relaxation time in presence of TAFF, $\epsilon = 1/I_0^2(\nu)$ is the flux creep factor, $I_p$ is the modified Bessel's function of the first kind of order $p$, and $\nu = U/2k_BT$; $U$ being the pinning barrier relevant for the thermally activated vortex motion. Furthermore, the normal fluid skin depth, $\delta_{nf}$ introduces an additional correction such that eqn. (3) is modified to the following:

$$\lambda^2_{eff,CC} = (\lambda_L^2 - i\frac{\rho_v^{TAFF}}{\mu_0\omega})/(1 + 2i\frac{\lambda_L^2}{\delta_{nf}^2}), \qquad (5)$$

$\delta_{nf}$ follows the phenomenological variation[26] of the form $\delta_{nf}^2(t,h) = \frac{\left(\frac{2\rho_n}{\mu_0\omega}\right)}{1-f(t,h)}$ which is complementary to the $\lambda_L$ variation considering they come from normal and superconducting electrons respectively in the framework of the two-fluid model: $\lambda_L^2(t,h) = \frac{\lambda_L^2}{f(t,h)}$ where $f(t,h) = (1-t^4)(1-h)$ with $t = T/T_c$ and $h = H/H_{c2}$; $T_c$ and $H_{c2}$ being the zero-field superconducting transition temperature and upper critical field respectively.

## 3. Sample preparation and experimental details

### 3.1 *Sample growth*

Our sample consists of a 20-nm-thick (d) amorphous Molybdenum Germanium (*a*-MoGe) thin film, grown on (100) oriented MgO substrate by pulsed laser deposition (PLD) technique, with $T_c \sim 7K$ (similar to the samples in ref. 30, 31). An arc-melted $Mo_{70}Ge_{30}$ target was ablated using a 248 nm excimer laser to deposit the amorphous thin film, keeping the substrate at room temperature. The sample was capped with a 2-nm-thick Si layer to prevent surface oxidation. To maintain the stoichiometry of the film close to the stoichiometry of the target, laser pulses of comparatively high energy density, ~240 $mJ/mm^2$ per pulse with a repetition rate of 10 Hz, were bombarded. The growth rate was ~ 1 nm/100 pulses. For two-coil penetration depth measurements, the sample was deposited in the form of an 8-mm-diameter disk using a shadow mask (inset of Fig. 1(c)). In earlier measurements[30], the field-cooled and zero-field-cooled ac susceptibility response performed on similar samples in the vortex state showed that vortex pinning in these samples is very weak: The field-cooled and zero-field-cooled ac susceptibility response in the vortex state overlap entirely with each other at fields above $500\ Oe$.

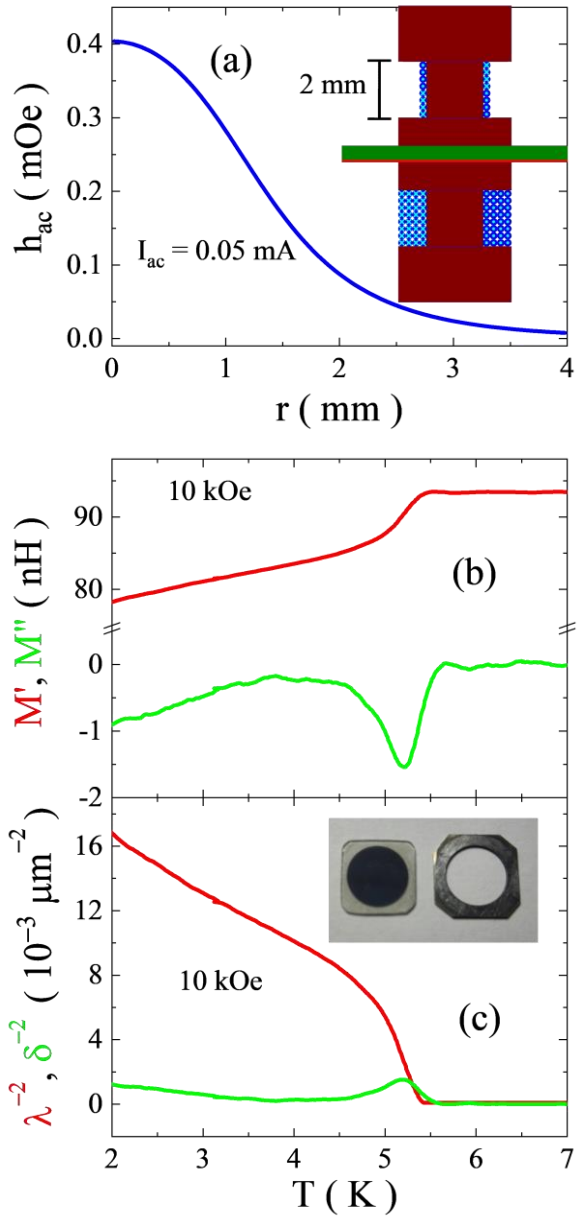

*Figure 1. (a) Radial distribution of ac field amplitude generated across the sample plane for passing 0.05 mA ac excitation through primary quadrupole coil; (inset) Schematic of two-coil setup: quadrupole as drive coil (top) and dipole as pick-up coil (bottom) with sample sandwiched in between. The coil wire diameter (50 μm) is drawn bigger than the actual for clarity. (b) Temperature dependence of M' and M'' at 10 kOe. (c) Corresponding $\lambda^{-2}$ and $\delta^{-2}$ as a function of temperature at 10 kOe; (inset) 8-mm-diameter a-MoGe sample grown on MgO substrate (left) with the mask (right).*

### 3.2 Penetration depth from Two-coil mutual inductance measurements

In the two-coil setup[16, 17, 18, 20], the 8-mm-diameter superconducting thin film is sandwiched between a miniature quadrupole primary and secondary dipole coil (inset of Fig. 1(a)) and placed inside a $^4$He cryostat fitted with a superconducting solenoid. Both primary and secondary coils are wound on bobbins with 2 mm diameters made out of Delrin. The quadrupolar primary coil is made by winding 15 turns clockwise in the one half of the coil and 15 turns anticlockwise in the other half of the coil. The dipolar secondary consists of 120 turns wound in 4 layers. 50 μm diameter copper wires are used for both coils. A small ac current, $I_{ac}$, with amplitude 0.05 mA, 30 kHz is passed through primary and the resulting in-phase and out-of-phase components of the induced voltage, $V^{in}$ and $V^{out}$, in the secondary is measured using a lock-in amplifier. The complex mutual inductance ($M = M' + iM''$) between the two coils is given by: $M'(M'') = V^{out}(V^{in})/\omega I_{ac}$. The use of a quadrupolar coil as the primary ensures fast radial decay of the ac magnetic field on the film (Fig. 1 (a)), which drops to about 2% of the central value at the edge of the film. This minimizes edge effects such as surface[32, 33, 34] and geometric barriers[35, 36] from having significant effects on the measured shielding response. In order to extract $\lambda_{eff}^2$ from $M$, we first note that $\lambda_{eff}$ can be represented as: $\lambda_{eff}^{-2} = \lambda^{-2} + i\delta^{-2}$ where $\lambda^{-2}$ is the inductive response whereas $\delta^{-2}$ is the dissipative response (skin depth). However, extracting $\lambda_{eff}^{-2}$ from $M$ is not straightforward since the $M'$ and $M''$ both depend on $\lambda$ and $\delta$. For this, we numerically solve the coupled Maxwell and London equations for the geometry of our coils and sample using finite element analysis and create a lookup table of $M = M' + iM''$ for a range of values of $\lambda$ and $\delta$. Details of this procedure can be found in ref. 37, 17 and 18. $\lambda_{eff}$ is then extracted by comparing the experimentally measured value of $M$ with the corresponding value in the lookup table. In Fig. 1 (b) and (c), we show representative plots of the temperature variation of $M (= M' + iM'')$ and corresponding $\lambda^{-2}$ and $\delta^{-2}$ at $10\ kOe$ respectively.

## 4. Results

In Fig. 2 (a) and (b), we plot the temperature variation of $\lambda^{-2}$ and $\delta^{-2}$ for different fields. First, we concentrate on $\lambda^{-2}$. We observe that $\lambda^{-2}(T \to 0)$ decreases as a function of field signifying an increase in the Campbell penetration depth. With increase in temperature, $\lambda^{-2}$ smoothly decreases at low temperatures before it drops rapidly eventually going below the resolution limit. From earlier measurements, we know that the vortex state in 20-nm-thick $a$-MoGe film undergoes two transitions: From a vortex solid (VS) at low fields and temperatures to a hexatic vortex fluid (HVF), and then from a hexatic vortex fluid (HVF) to an isotropic vortex liquid (IVL). In Fig. 2 (c), we plot the temperature ($T^*$) for every field at which $\lambda^{-2}(T > T^*)$ drops below our resolution limit of $10^8\ m^{-2}$. In the same graph we plot the locus of the transitions from VS to HVF and HVF to IVL obtained earlier from scanning tunnelling spectroscopy and magneto-transport measurements[30]. At low temperatures, $T^*(H)$ is just below the HVF-IVL boundary[30] showing that the screening response vanishes in the IVL. However, we do not see any signature of the VS-HVF transition in the temperature variation of $\lambda^{-2}$. This is due to the fact that at low temperature the mobility of the vortices in the HVF is extremely slow[31], and the ac pinning response is practically indistinguishable from the vortex solid. This is also consistent with earlier magneto-transport measurements where no sharp change in pinning properties was observed across the VS-HVF transition at low temperatures. In fact, it was shown earlier[31] that the Larkin-Ovchinnikov[38] collective pinning model, originally developed to explain vortex pinning in an imperfect vortex solid is largely applicable in the HVF state as well, at temperatures well below $T_c$. At higher temperatures $T^*$ bends towards the VS-HVF boundary. This can be understood from the fact that the mobility of the vortices increases with increasing temperatures due to thermal fluctuations, and now even in the HVF the vortices experience very little restoring force due to pinning, thereby destroying the shielding response.

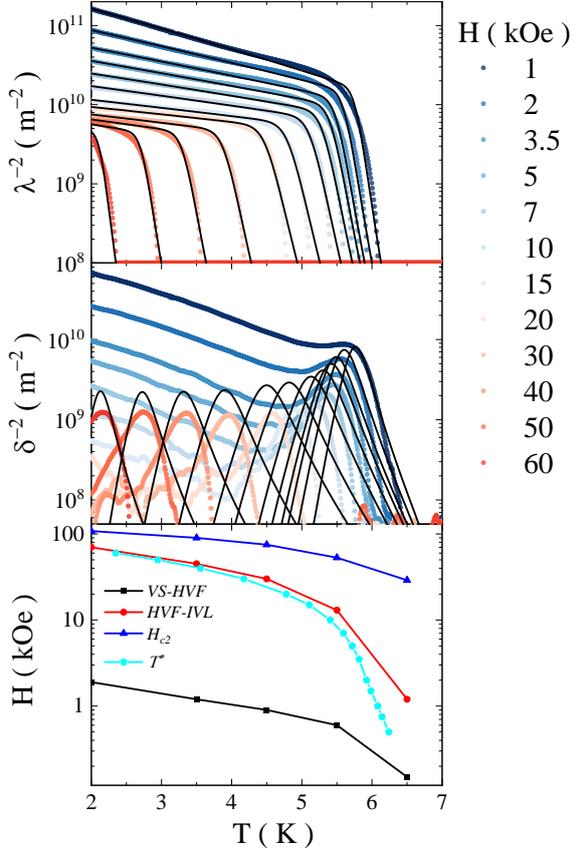

Figure 2. (a), (b) Temperature variation of experimentally measured $\lambda^{-2}$ and $\delta^{-2}$ respectively at the different fields (1 - 60 kOe), along with the fit to eqn. (6) with the best-fit parameters following the method discussed in appendix. (c) "$\lambda^{-2}$-vanishing" temperature points ($T^*$) were noted for all the fields from Fig. 2 (a) where $\lambda^{-2}(T > T^*)$ drops below the resolution limit $10^8\ m^{-2}$ and plotted (Cyan connected pentagons) in the H-T parameter space along with earlier phase space boundaries taken from ref. 30: VS-to-HVF (connected black squares), HVF-to-IVL (connected red circles) and $H_{c2}$ (connected blue triangles) respectively (VS: Vortex solid, HVF: Hexatic vortex fluid and IVL: Isotropic vortex liquid).

Coming to $\delta^{-2}$ (Fig. 2(b)) we observe that for fields above 10 kOe, $\delta^{-2}$ shows a single dissipative peak close to $T^*$ as expected from fluctuations close to a transition. However, at low fields, in addition to this dissipative peak, we observe an increase in $\delta^{-2}$ with decrease in temperatures. This increase in $\delta^{-2}$ at low temperatures indicates the presence of additional dissipative modes at low temperatures which is not accounted for by the theoretical framework used here. We will comment more on that later.

## 5. Discussion

In order to quantitatively analyse the data, we fit the data with eqn. (5). First, we note that in order for the harmonic approximation (used to derive eqn. (5)) to hold, the ac excitation needs to satisfy the following condition[8]: $h_{ac} \ll \mu_0 J_c \lambda_C = (\mu_0 J_c B r_f)^{1/2} \equiv h_p$. Using previously published data[31] on magneto-transport properties of a similar film, $J_c B \sim 2 \times 10^7 - 3 \times 10^8 N/m^3$ within our field of interest ($1 - 60$ $kOe$) and $r_f \sim \xi \sim 5.5$ $nm$, where $\xi$ is the Ginzburg-Landau (GL) coherence length. Using these values we estimate $h_p \sim 3 - 14$ $Oe$ which is several order of magnitudes larger than our $h_{ac}$ (Fig. 1(a)). Furthermore, from the normal state resistivity of the film, $\rho_N \sim 1.55$ $\mu\Omega - m$ we estimate $\delta_{nf}(0) \sim 3.5$ $mm$. Since $\lambda_L(0) \sim 587$ $nm$, we obtain $\frac{2\lambda_L^2}{\delta_{nf}^2} \sim 5 \times 10^{-8}$. We can therefore neglect this term in eqn. (5) and set the denominator to unity. The simplified equation now reads as:

$$\lambda^{-2} + i\delta^{-2} = (\lambda_L^2 - i\frac{\rho_{ff}}{\mu_0 \omega}\frac{\epsilon + i\omega\tau}{1 + i\omega\tau})^{-1} \qquad (6)$$

We now attempt to fit $\lambda^{-2}$ with the real part of the right-hand side of eqn. (6). However, in order to fit the temperature dependence we need to make some assumptions on the temperature variation of $\alpha_L, \eta$ $and$ $U$. The temperature dependence of $\eta$ is assumed to follow the variation of $H_{c2}$ according to the Bardeen-Stephen formula[22, 23]: $\eta = \frac{\mu_0 \phi_0 H_{c2}}{\rho_n}$. $H_{c2}$ was earlier measured[30] from isotherms in magneto-transport experiments and defined from the criterion, $\rho(H_{c2}) \sim 0.95\rho_n$. As shown in the inset of Fig. 3, $H_{c2}$ follows the empirical variation[30]: $H_{c2} \propto \left(\frac{1-t^2}{1+t^2}\right)^{0.66}$. Therefore, variation of $\eta$ is taken as:

$\eta = \eta_0 \left(\frac{1-t^2}{1+t^2}\right)^{0.66}$, where $\eta_0 = \frac{\mu_0 \phi_0 H_{c2}(0)}{\rho_n}$. Our sample (amorphous MoGe thin film) being an s-wave superconductor, $\eta_0$ is assumed to be field-independent[39]. For the variation of $\alpha_L$ and $U$, several different functional forms have been assumed in literature. In the framework of 2D collective pinning model, the activation barrier $U$ governing the thermal creep rate can be estimated by the shearing elastic energy of the vortex lattice due to the small displacement of a single vortex as[40, 41]: $U \sim C_{66} \left(\frac{\xi}{R_c}\right)^2 V_c$, where the displacement $u$ has been taken as the core dimension, $\xi$, and $C_{66}$, $R_c$ and $V_c$ are the shear modulus, transverse Larkin length, and collective pinning volume of the VL respectively. In the case of the 2D collective pinning model, $V_c \sim R_c^2 d$ and hence $U \sim C_{66} \xi^2 d$. Since[38] $C_{66} \sim \mu_0 H_c^2 h(1-h)^2$, $U(t) \propto H_c^2(t)\xi^2(t)$. Using the known empirical dependences[11, 42], $H_c(t) \sim (1-t^2)$ and $\xi(t) \sim \left(\frac{1+t^2}{1-t^2}\right)^{1/2}$ one obtains,

$$U(t, H) = U_0(H)(1-t^2)(1+t^2) \tag{7}$$

The restoring force term $\alpha_L \vec{u}$ can be represented as the derivative of pinning potential $U$, which means that the spring constant $\alpha_L$ can be expressed as $\frac{\partial^2 U}{\partial \vec{u}^2}$, i.e. $\alpha_L$ depends on the curvature of the pinning potential at the bottom. Though the exact shape and curvature of $U$ are difficult to predict, an estimate of $\alpha_L$ can be found from the energy consideration of the core of the vortex when subjected to the small displacement considered in eqn. (1). If we equate the total condensation energy in the displaced core with the stored elastic energy: $\mu_0 H_c^2 (\pi \xi^2 d) = \frac{1}{2}\alpha_L \xi^2 d$, we get the temperature variation of $\alpha_L$ as[43, 44]:

$$\alpha_L(t) \sim H_c^2(t) \sim (1-t^2)^2 \tag{8}$$

On the other hand, a significant modification of this functional form was suggested by Feigel'man et al.[45] and Koshelev et al.[46] considering the effect of thermal fluctuations. They suggested that the smearing of the pinning potential due to thermal fluctuations would result in the exponential decay of $\alpha_L$ and $U$. Experimentally, such exponential decay was indeed observed for YBa$_2$Cu$_3$O$_7$ and Bi$_2$Sr$_2$CaCu$_2$O$_8$ thin films[21, 47, 48] for temperature lower than $0.8T_c$. Since the 2-dimensional vortex state in thin $a$-MoGe films is extremely sensitive to small perturbations[49], it is expected that the vortex lattice would also be susceptible to thermal fluctuations. Indeed, our attempts to fit the data with the temperature dependences in eqns. (7) and (8) resulted in poor fit particularly at low magnetic fields (Fig. 3). On the other hand, much better fit over the entire field range was obtained (Fig. 3 and Fig. 2(a)) by assuming temperature dependence of the form: $\alpha_L \sim \alpha_{L0}\ e^{-\left(\frac{T}{T_0}\right)}$ and $U \sim U_0\ e^{-\left(\frac{T}{T_0}\right)}$ (See Appendix for the best fit procedure adopted in this work). Nevertheless, a small deviation is observed at higher temperatures close to the knee region above which $\lambda^{-2}$ drops rapidly. This is however unsurprising since this is close to the boundary where the harmonic approximation assumed in eqn. (5) will start to break down. For 50 and 60 kOe the temperature range of data is too small to perform a reliable best fit, but the qualitative variation is captured by using extrapolated parameters from lower fields.

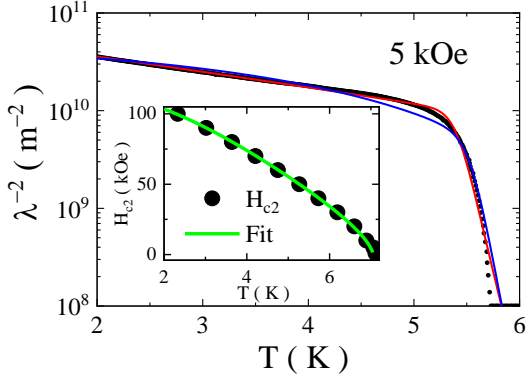

Figure 3. Experimental $\lambda^{-2}$ data at a representative field (5 kOe) (black circles), along with the fit to eqn. (6) using (i) $\alpha_L = \alpha_{L0}e^{-T/T_0}$, $U = U_0 e^{-T/T_0}$ (red line) and (ii) $\alpha_L = \alpha_{L0}(1-t^2)^2$, $U = U_0(1-t^2)(1+t^2)$ (blue line). (inset) Temperature dependence of $H_{c2}$ (kOe) from ref. 30 (black circles), defined by the criterion: $\rho(H_{c2}) \sim 0.95\rho_n$, along with the fit: $H_{c2}(0)\left[\frac{1-t^2}{1+t^2}\right]^{0.66}$ (green line), with $H_{c2}(0) = 115\ kOe, t = \frac{T}{T_c}, T_c = 7.05\ K$.

Fig. 4 show the best fit parameters extracted from these fits. In Fig. 4(a), $U_0$ shows a monotonic decrease that can be fitted with a simple power law with exponent $H^{-0.82}$. This is consistent with earlier studies[50] on *a*-MoGe films even though our exponent is somewhat larger than 0.66 reported in ref. 50. On the other hand, $T_0$ and $\alpha_{L0}$ exhibit non-monotonic behaviour (Fig. 4(b) and 4(c) respectively). $\alpha_{L0}$ initially decreases with field and exhibits a shallow minimum at 15 kOe. This minimum is qualitatively similar to that observed in the variation of $J_c$ with field measured on similar samples[31] and can be understood based on the theory of collective pinning. Within Larkin-Ovchinnikov[38] theory of collective pinning, the pinning force on a vortex (per unit length) for a displacement *u* is given by,

$$F_p^{\phi_0} = \alpha_{L0}u \approx \frac{\phi_0}{B}\frac{\left(\langle(f_{\alpha_{L0}})^2\rangle n_A\right)^{\frac{1}{2}}}{dR_c} \qquad (9)$$

where $f_{\alpha_{L0}}$ is the elementary restoring pinning force and $n_A$ is the areal density of pinning centres. $F_p^{\phi_0}$ is governed by two counteracting effects: The variation of $f_{\alpha_{L0}}$ and $\underline{R_c}$ with magnetic field. With increase in field, $f_{\alpha_{L0}}$ and $R_c$ control the behaviour of $\alpha_{L0}$ making it non-monotonic with field. Depending on the nature of pinning, the variation of $f_{\alpha_{L0}}$ follows the general form[51]: $\langle(f_{\alpha_{L0}})^2\rangle n_A \propto Ch^n(1-h)^2$, which suggests that: $F_p^{\phi_0} \propto h^{\frac{n}{2}-1}(1-h)/dR_c$, where $n = 1$ corresponds to pinning due to dislocation loops[52] and $n = 3$ due to impurities and vacancies[53] respectively. We have earlier observed that $n = 1$ correctly describes the pinning in similar sample[31]. Thus, at low fields the initial decrease in $\alpha_{L0}$ comes from the dominant numerator because of decrease in elementary pinning force with increasing field. However, above a certain field the decrease in $R_c$ [31] with increasing fields starts to dominate, thus increasing $\alpha_{L0}$ (Fig. 4 (c)). Similar field variation in $J_c$ from the magneto-resistance measurements[31] was observed on a similar sample (inset of Fig. 4 (c)) as a precursor of the peak effect observed in the sample. We could not capture the peak at higher fields since the shielding response goes below our resolution limit. To further analyse the data quantitatively, we express eqn. (9) in terms of pinning force density ($F_p$):

$$F_p \approx \alpha_{L0} \xi \left(\frac{B}{\phi_0}\right) \propto \frac{h^{\frac{n}{2}}(1-h)}{dR_c} \tag{10}$$

where displacement of a vortex is assumed to be of the order of coherence length $\xi$. From field variation of $R_c$ obtained from scanning tunnelling spectroscopic images in ref. 31, in Fig. 4(d), we also plot the variation of $F_p$ as a function of normalized field $h (= H/H_{c2})$ from eqn. (10) where the proportionality constant is taken as an adjustable parameter. The qualitative agreement is indeed very good when we consider that $R_c$ has been measured on a different sample. Finally, we observe that $T_0$ increases from 1.5 – 5 K up to 20 kOe, and then exhibits a gentle decrease in the same range of fields where we observe the increase in $\alpha_{L0}$. Even though we do not have a model to explain this variation at the moment, we believe that the initial increase is related to the increase in rigidity of the VL with field as the vortex lattice is squeezed. The decrease in high field is more difficult to understand. However, in this range the error bar on the extracted value of $T_0$ is large (due to the lower temperature range of the fit) and it is difficult at the moment to assess the significance of this decrease. The variation of $T_0$ with field needs to be explored further in future.

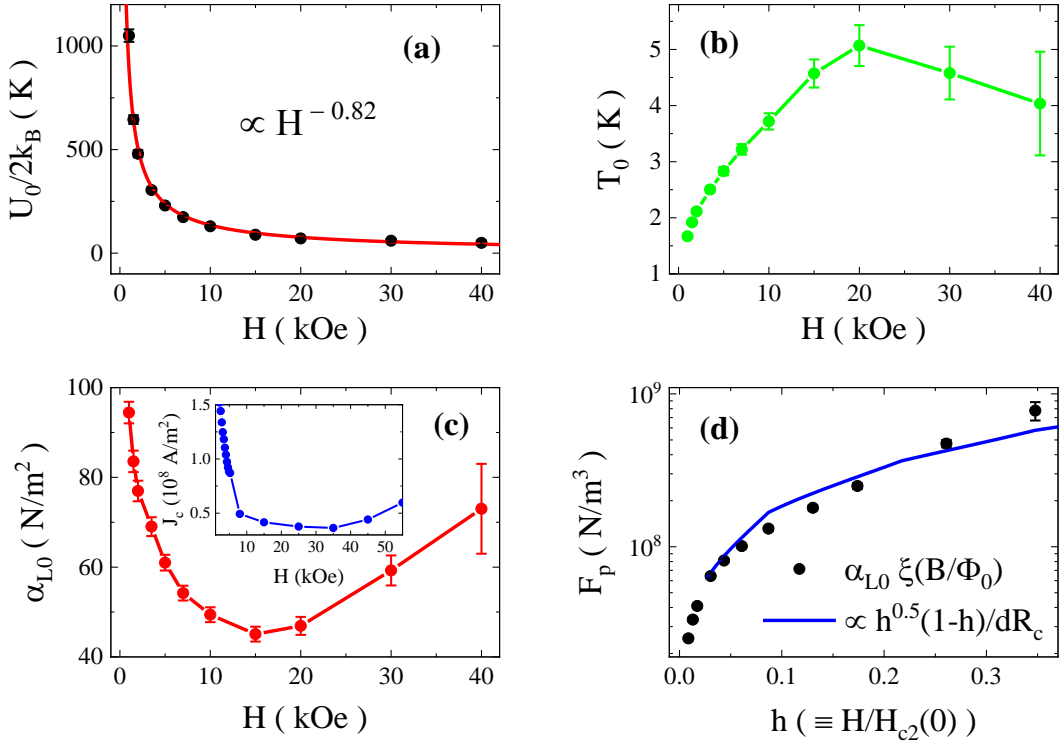

*Figure 4. (a) $U_0/2k_B$ vs H (black circles) which follows power law decay (red line): $U_0 \propto H^{-0.82}$. (b) Field variation of the characteristic temperature $T_0$ (Used in the the temperature dependence: $\alpha_L(U) = \alpha_{L0}(U_0)e^{-T/T_0}$) (connected green circles). (c) Field dependence of $\alpha_{L0}$ (connected red circles). (inset) Critical current density ($J_c$) as a function of field (blue connected circles) measured on a similar sample from Ref. 31. (d) Pinning force density $F_p (= \alpha_{L0}\xi B/\phi_0)$ vs normalized field $h (= H/H_{c2}(0))$ (black circles) along with the theoretical field variations proportional to $h^{0.5}(1-h)/dR_c$ (blue line) (semi-log scale). Field variation of $R_c$ is taken from ref. 31 and $d = 20$ nm is the film thickness. Error bars for the parameters were determined following the protocol explained in appendix.*

We also looked at the imaginary part of eqn. (6) which corresponds to $\delta^{-2}$. In Fig. 2 (b) we plot the imaginary part of the right-hand side of eqn. (6) using the same parameters as those used to fit $\lambda^{-2}$. At high fields the simulated curve qualitatively captures the dissipation peak observed close to $T^*$. However, at low fields we observe an additional increase at low temperatures which is not captured within this model. This increase signals some additional mode of dissipation present in the system. One possibility is that at low fields the oscillatory field excites additional internal modes within a Larkin domain in the soft lattice that is not accounted for in the mean-field description. However, this intriguing pronounced feature is at present unexplained and would form the basis of future studies.

## 6. Conclusion

In this paper, we have demonstrated a method of extracting vortex parameters from penetration depth measurements using the low-frequency mutual inductance technique. We fitted the measured $\lambda^{-2}$ $vs$ $T$ plots at different fields with model developed by Coffey and Clem[26], where the vortex parameters were determined as fitted parameters. The accuracy of the extracted parameter values depends on their correct temperature dependence models. Though the CC model has been extensively used (with some approximations) to extract vortex parameters from the microwave vortex resistivity measurements[21, 29, 54], it has not been used very much for low-frequency ac susceptibility measurements barring a few[55, 56]. Our data fitted well for exponential temperature variations of both Labusch parameter and activation potential barrier, suggesting that the dominant effect of temperature comes from the smearing of pinning potential due to thermal fluctuations. However, using the CC model we could not fully capture the variation of the skin depth signifying loss in the system which suggests that there might be additional modes of dissipation present in the system beyond the model presented here. One limitation of our approach is that we have to pre-assume the temperature dependence of some parameters in our analysis. If the chosen temperature dependences are not accurate for a material, the shape of the fitted curve does not reproduce the experimental data accurately. If one knows the precise temperature dependence of the vortex parameters for the concerned material, the estimate of the final result only gets better.

**Author contributions**

SM and SB performed the measurements and analysed the data. SB, SM, JJ and VB prepared the sample and performed preliminary characterisation. PR conceived the problem and supervised the project. SM and PR wrote the paper with inputs from all the authors.

**Appendix: Fitting procedure and error estimate of the vortex parameters:**

We start with the eqn. (6) which was used to fit the experimental $\lambda^{-2}$ data, where $\tau = \tau_0 \frac{I_0^2(\nu)-1}{I_1(\nu)I_0(\nu)}$ and $\epsilon = \frac{1}{I_0^2(\nu)}$ depend on the parameters: $\tau_0 = \frac{\eta}{\alpha_L}$ and $\nu = \frac{U}{2k_BT}$. We have used a fixed variation of $\eta$ calculated from $\frac{\mu_0 \phi_0 H_{c2}}{\rho_n}$, while the remaining free parameters which were varied to fit eqn. (6) to $\lambda^{-2}$ data are $\alpha_L$ and $U$. We have taken the temperature dependence as: $\alpha_L \sim \alpha_{L0} \, e^{-\left(\frac{T}{T_0}\right)}$ and $U \sim U_0 \, e^{-\left(\frac{T}{T_0}\right)}$, where the characteristic temperature $T_0$ is the same for both $\alpha_L$ and $U$ (see section 5). Thus, we have three free parameters which can be varied to find the best fit to eqn. (6): $T_0, \alpha_{L0}$ and $U_0/2k_B$, which were plotted as a function of field in Fig. 4.

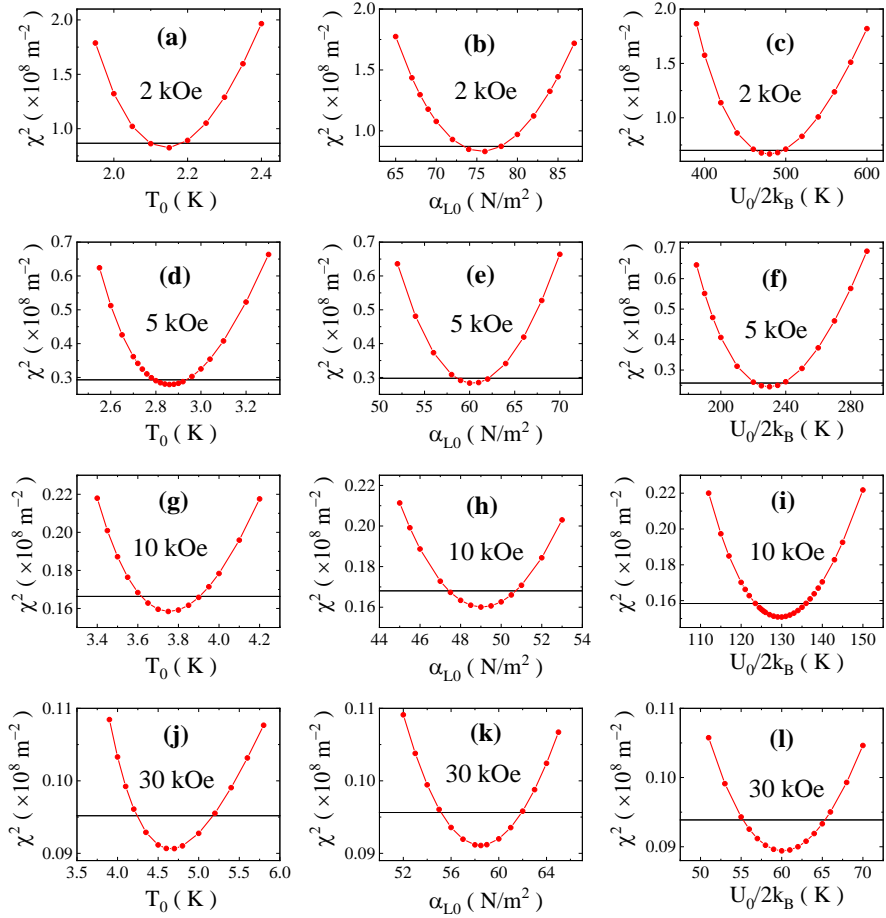

*Figure 5. $\chi^2$ as a function of $T_0, \alpha_{L0}$ and $U_0/2k_B$ respectively for four representative fields: (a)-(c) 2 kOe; (d)-(f) 5 kOe; (g)-(i) 10 kOe and (j)-(l) 30 kOe. The horizontal black line in each of the plots is at 1.05 times of the minimum $\chi^2$ value: spread of parameter values below the line has been chosen as the error bar of that parameter for the given field.*

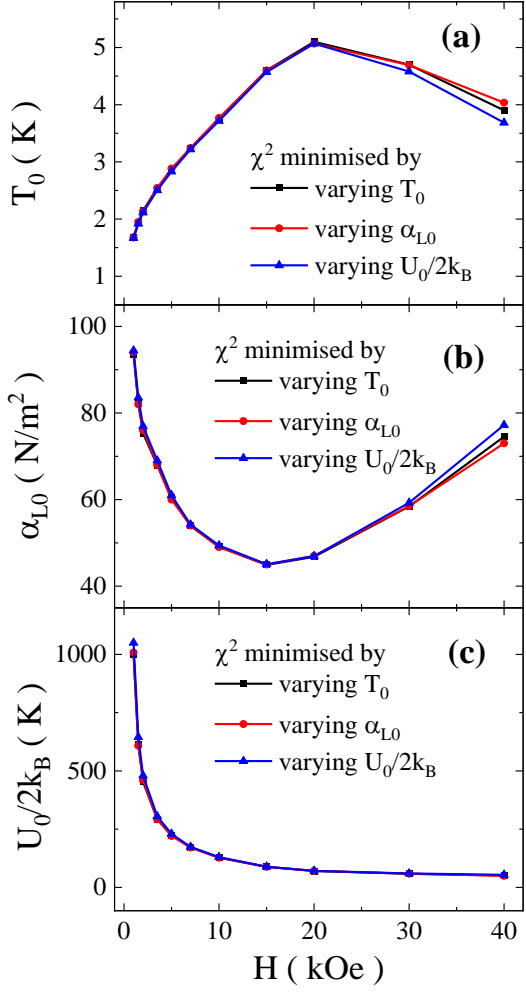

*Figure 6. (a)-(c) Field dependence of $T_0$, $\alpha_{L0}$ and $U_0/2k_B$ values at the $\chi^2$ minima points respectively. In each plot, minimisation was done by varying each parameter individually: minima of the $\chi^2$ vs $T_0$ plots are given by connected black squares, minima of the $\chi^2$ vs $\alpha_{L0}$ plots by connected red circles and minima of the $\chi^2$ vs $U_0/2k_B$ plots by connected blue triangles.*

Goodness of fit is calculated by the formula: $\chi^2 = \frac{1}{N}\Sigma\frac{(y-y_{fit})^2}{y_{fit}}$, where y is the experimental $\lambda^{-2}$ data in Fig. 2(a) and $y_{fit}$ is calculated using right-hand side of eqn. (6); the normalisation factor 1/N was to account for the fact that different datasets had different number of points, N. Here the free parameters are $\alpha_{L0}, T_0$ and $U_0/2k_B$. To carry out the error analysis we used the following protocol: One particular parameter was fixed and the other two were varied freely so as to have the best fit using the FindFit function in Mathematica, and the corresponding $\chi^2$ was calculated. In this procedure the best fit was given by the set of values where $\chi^2$ is minimum. The results are shown in Fig. 5 (a)-(l) for 4 representative fields (*2, 5, 10 and 30 kOe*). In principle, all three curves should give the same value of $\chi^2$ at the minimum. The slight difference is owing to the fact that the in-built FindFit function minimises the least-square error instead of $\chi^2$. Nevertheless, the extracted parameter values from the minima of either of these three curves are very close to each other and shown in Fig. 6. The set of parameters ($\alpha_{L0}, T_0 \text{ and } U_0/2k_B$) corresponding to the lowest of the three minima are plotted in Fig. 4. Here we did not include parameters for *50 and 60 kOe* since the temperature range of data is too small to perform a reliable best fit. Nevertheless, the qualitative variation at these fields is still captured by using extrapolated parameters from lower fields.

The variation of $\chi^2$ for each of the three free parameters, allows us to estimate of the error bar. The set of parameter values that do not increase $\chi^2$ beyond 5% of the minimum $\chi^2$ value (below the horizontal black line in each plot in Fig. 5), is taken as the error bar for that particular parameter.